\documentclass[twocolumn]{aastex63}
\usepackage{amsmath}
\usepackage[T1]{fontenc}
\usepackage{upgreek}

\newcommand{\PSUAA}{Department of Astronomy \& Astrophysics, 525 Davey Laboratory, The Pennsylvania State University, University Park, PA, 16802, USA}
\newcommand{\PSUCEHW}{Center for Exoplanets and Habitable Worlds, 525 Davey Laboratory, The Pennsylvania State University, University Park, PA, 16802, USA}
\newcommand{\PSETI}{Penn State Extraterrestrial Intelligence Center, 525 Davey Laboratory, The Pennsylvania State University, University Park, PA, 16802, USA}
\newcommand{\UA}{Steward Observatory, The University of Arizona, 933 N.\ Cherry Ave, Tucson, AZ 85721, USA}
\newcommand{\Penn}{Department of Physics and Astronomy, University of Pennsylvania, 209 S 33rd St, Philadelphia, PA 19104, USA}

\newcommand{\GoddardESAL}{Exoplanets and Stellar Astrophysics Laboratory, NASA Goddard Space Flight Center, Greenbelt, MD 20771, USA}

\newcommand{\NOAO}{NSF's National Optical-Infrared Astronomy Research Laboratory, 950 N.\ Cherry Ave., Tucson, AZ 85719, USA}

\newcommand{\Macquarie}{Department of Physics and Astronomy, Macquarie University, Balaclava Road, North Ryde, NSW 2109, Australia }

\newcommand{\JPL}{Jet Propulsion Laboratory, California Institute of Technology, 4800 Oak Grove Drive, Pasadena, California 91109}
\newcommand{\MIT}{Kavli Institute for Astrophysics and Space Research, Massachusetts Institute of Technology, Cambridge, MA, USA}
\newcommand{\UCI}{Department of Physics \& Astronomy, The University of California, Irvine, Irvine, CA 92697, USA}
\newcommand{\Carleton}{Carleton College, One North College St., Northfield, MN 55057, USA}
\newcommand{\PSUICS}{Institute for CyberScience, The Pennsylvania State University, University Park, PA, 16802, USA}

\newcommand{\Princeton}{Department of Astrophysical Sciences, Princeton University, 4 Ivy Lane, Princeton, NJ 08540, USA}
\newcommand{\RUSSELL}{Henry Norris Russell Fellow}
\newcommand{\STScI}{Space Telescope Science Institute, 3700 San Martin Dr, Baltimore, MD 21218, USA}

\newcommand{\cms}{cm s$^{-1}$}
\newcommand{\teff}{T_{\rm eff}}

\received{September 10, 2020}
\revised{November 17, 2020}
\accepted{December 29, 2020}
\submitjournal{AJ}

\shorttitle{NEID Earth Twin Survey}
\shortauthors{Gupta et al.}

\graphicspath{{./}{figures/}}

\begin{document}

\title{Target Prioritization and Observing Strategies for the NEID Earth Twin Survey}

\correspondingauthor{Arvind F.\ Gupta}
\email{arvind@psu.edu}

\author[0000-0002-5463-9980]{Arvind F.\ Gupta}
\affil{\PSUAA}
\affil{\PSUCEHW}

\author[0000-0001-6160-5888]{Jason T.\ Wright}
\altaffiliation{Instrument Team Project Scientist}
\affil{\PSUAA}
\affil{\PSUCEHW}
\affil{\PSETI}

\author[0000-0003-0149-9678]{Paul Robertson}
\altaffiliation{Instrument Team Project Scientist}
\affil{\UCI}

\author[0000-0003-1312-9391]{Samuel Halverson}
\altaffiliation{Sagan Fellow}
\affil{\JPL}
\affil{\MIT}

\author[0000-0002-4927-9925]{Jacob Luhn}
\affil{\PSUAA}
\affil{\PSUCEHW}

\author[0000-0001-8127-5775]{Arpita Roy}
\affil{\STScI}

\author[0000-0001-9596-7983]{Suvrath Mahadevan}
\altaffiliation{Principal Investigator}
\affil{\PSUAA}
\affil{\PSUCEHW}

\author[0000-0001-6545-639X]{Eric B.\ Ford}
\affil{\PSUAA}
\affil{\PSUCEHW}
\affil{\PSUICS}


\author[0000-0003-4384-7220]{Chad F.\ Bender}
\affil{\UA}

\author[0000-0002-6096-1749]{Cullen H.\ Blake}
\affil{\Penn}

\author[0000-0002-1664-3102]{Fred Hearty}
\affil{\PSUAA}
\affil{\PSUCEHW}

\author[0000-0001-8401-4300]{Shubham Kanodia}
\affil{\PSUAA}
\affil{\PSUCEHW}

\author[0000-0002-9632-9382]{Sarah E.\ Logsdon}
\affil{\NOAO}

\author[0000-0003-0241-8956]{Michael W.\ McElwain}
\affil{\GoddardESAL}

\author{Andrew Monson}
\affil{\PSUAA}
\affil{\PSUCEHW}

\author[0000-0001-8720-5612]{Joe P.\ Ninan}
\affil{\PSUAA}
\affil{\PSUCEHW}

\author[0000-0002-4046-987X]{Christian Schwab}
\affil{\Macquarie}

\author[0000-0001-7409-5688]{Gu{\dh}mundur Stef\'ansson}
\altaffiliation{\RUSSELL}
\affil{\Princeton}

\author[0000-0002-4788-8858]{Ryan C.\ Terrien}
\affil{\Carleton}


\begin{abstract}
NEID is a high-resolution optical spectrograph on the WIYN 3.5-m telescope at Kitt Peak National Observatory and will soon join the new generation of extreme precision radial velocity instruments in operation around the world. We plan to use the instrument to conduct the NEID Earth Twin Survey (NETS) over the course of the next 5 years, collecting hundreds of observations of some of the nearest and brightest stars in an effort to probe the regime of Earth-mass exoplanets. Even if we take advantage of the extreme instrumental precision conferred by NEID, it will remain difficult to disentangle the weak ($\sim$10 \cms) signals induced by such low-mass, long-period exoplanets from stellar noise for all but the quietest host stars. In this work, we present a set of quantitative selection metrics which we use to identify an initial NETS target list consisting of stars conducive to the detection of exoplanets in the regime of interest. We also outline a set of observing strategies with which we aim to mitigate uncertainty contributions from intrinsic stellar variability and other sources of noise. 
\end{abstract}

\section{Introduction}

Following the discovery of 51 Pegasi b \citep{Mayor95}, the radial velocity (RV) technique reigned as the most prolific method of exoplanet detection before the focus shifted to transit discoveries once \textit{Kepler} \citep{Borucki10} revolutionized the field. But transit surveys are strongly biased towards the detection of exoplanets with short orbital periods and small separations from their host stars. This is particularly important in anticipation of future flagship mission concepts such as LUVOIR \citep{LUVOIR19} and HabEx \citep{Gaudi20}, both of which will be able to achieve direct imaging contrast levels sufficient for atmospheric biosignature characterization of Earth analogs around the $\sim150$ Sun-like stars within $15$ pc. While LUVOIR and HabEx will have the capability to conduct blind direct imaging searches to identify targets for subsequent characterization, the efficiency of these missions will be greatly improved if a target list can be defined ahead of time. The geometric transit probability for an exoplanet orbiting a Sun-like star at 1 AU is just $0.47\%$, so we expect to detect less than one low-inclination, transiting Earth analog among these nearby stars. \textit{Kepler}, \textit{TESS} \citep{Ricker15}, and future transit surveys will do little to directly inform the target lists of these flagship missions. RV measurements, on the other hand, are sensitive to exoplanets with a wider range of orbital inclinations. If we aim to study and characterize Earth-like exoplanets, blind RV surveys remain an essential tool.

NEID, a high-resolution (R$\sim$110,000) optical spectrograph (380 - 930 nm) \citep{Schwab16} recently installed on the WIYN 3.5-m telescope at Kitt Peak National Observatory, was built as part of the NASA-NSF Exoplanet Observational Research (NN-EXPLORE) initiative with the purpose of providing mass constraints for transiting exoplanet candidates down to the size of the Earth. As such, NEID was designed with an instrumental precision goal of 27 \cms\ \citep{Halverson16}. Once instrument commissioning is complete, it will join the ranks of extreme precision (sub-m s$^{-1}$) radial velocity (EPRV) spectrographs alongside EXPRES \citep{Jurgenson16}, ESPRESSO \citep{Pepe20}, HARPS \citep{Mayor03}, and HARPS-N \citep{Cosentino12}, among others.

In addition to providing RV follow-up support for transit surveys, NEID will be used to search for new exoplanets with time allocated through a guaranteed time observations (GTO) program. The NEID GTO program consists of 30 nights (300 hours) per year for 5 years beginning in early 2021; the majority of this allocation (80\%, or 240 hours per year) will be reserved for the NEID Earth Twin Survey (NETS), a search for Earth-mass exoplanets in the habitable zones of the closest and brightest G, K, and M dwarfs. NETS targets will be observed 50 times per year over the course of 5 years. To enhance our sensitivity to low-mass, long-period planets, we must leverage the high NEID instrumental precision by taking care to minimize other sources of RV uncertainty. In \S\ref{sec:noise}, we discuss the total RV error budget for NEID observations. We address all external sources and their expected contributions as well as potential calibration and mitigation strategies. We then identify an initial sample of NETS target candidates in \S\ref{sec:sample}, and in \S\ref{sec:selection} we outline a set of quantitative target prioritization metrics. We use these metrics to construct a Figure of Merit with which the potential targets are ranked and evaluated, and we discuss the selection of the final target list. Finally, in \S\ref{sec:observations}, we discuss observing guidelines that will improve the RV precision that NEID can achieve, both for NETS and for radial velocity exoplanet observations in general.

\section{Radial Velocity Precision}\label{sec:noise}
In the absence of external sources of noise, NEID is designed to achieve a single measurement precision of 27 \cms\ for bright stars \citep{Halverson16}.
In practice, however, contributions from photon noise, intrinsic stellar variability, and spectral contamination will elevate the achieved precision above the instrumental floor. The following sections discuss each of these sources as they apply to RV observations and we quantify their expected contributions to the total NEID single measurement precision.

\subsection{Photon Noise}
Photon noise places a fundamental limit on the achievable precision for the RV shift measured from a given spectrum. This precision is set by both the spectral signal to noise ratio (SNR) and the intrinsic stellar RV information content \citep{Bouchy01}. We calculate the expected photon noise for NEID observations using simulated 1-D spectra, which are derived from stellar models from the PHOENIX spectral library \citep{Husser13} and modulated by both the median measured atmospheric transmission at KPNO and a theoretical NEID/WIYN throughput model. The spectra are convolved with a stellar rotational broadening kernel ($v\sin{i}=2$ km s$^{-1}$) and instrument PSF model (Gaussian profile, $R$ $\sim$ 110,000), and binned to the spectral sampling of NEID (4.5 pixels FWHM). We then compute the weighted information content in each pixel \citep{Bouchy01, Murphy07}, including both photon and read noise, and combine all pixel weights to arrive at an integrated photon-limited RV uncertainty estimate for the full spectrum. Wavelengths with telluric features at $>$1\% absorption depth (and the surrounding spectral swaths at $\pm$15 km s$^{-1}$) are excluded from the uncertainty calculation, as these regions are currently ignored in the NEID RV pipeline. In addition to an integrated RV uncertainty, we compute order-by-order RV uncertainties and median spectral SNR values across the free spectral range of each echelle order. Finally, an empirical correction factor is applied to account for the small loss of information content expected due to the NEID cross-correlation function (CCF) binary mask having a finite width, which dilutes the effective information content when measuring RVs via the CCF technique.

\begin{figure*}
    \centering
    \epsscale{1.0}
    \plotone{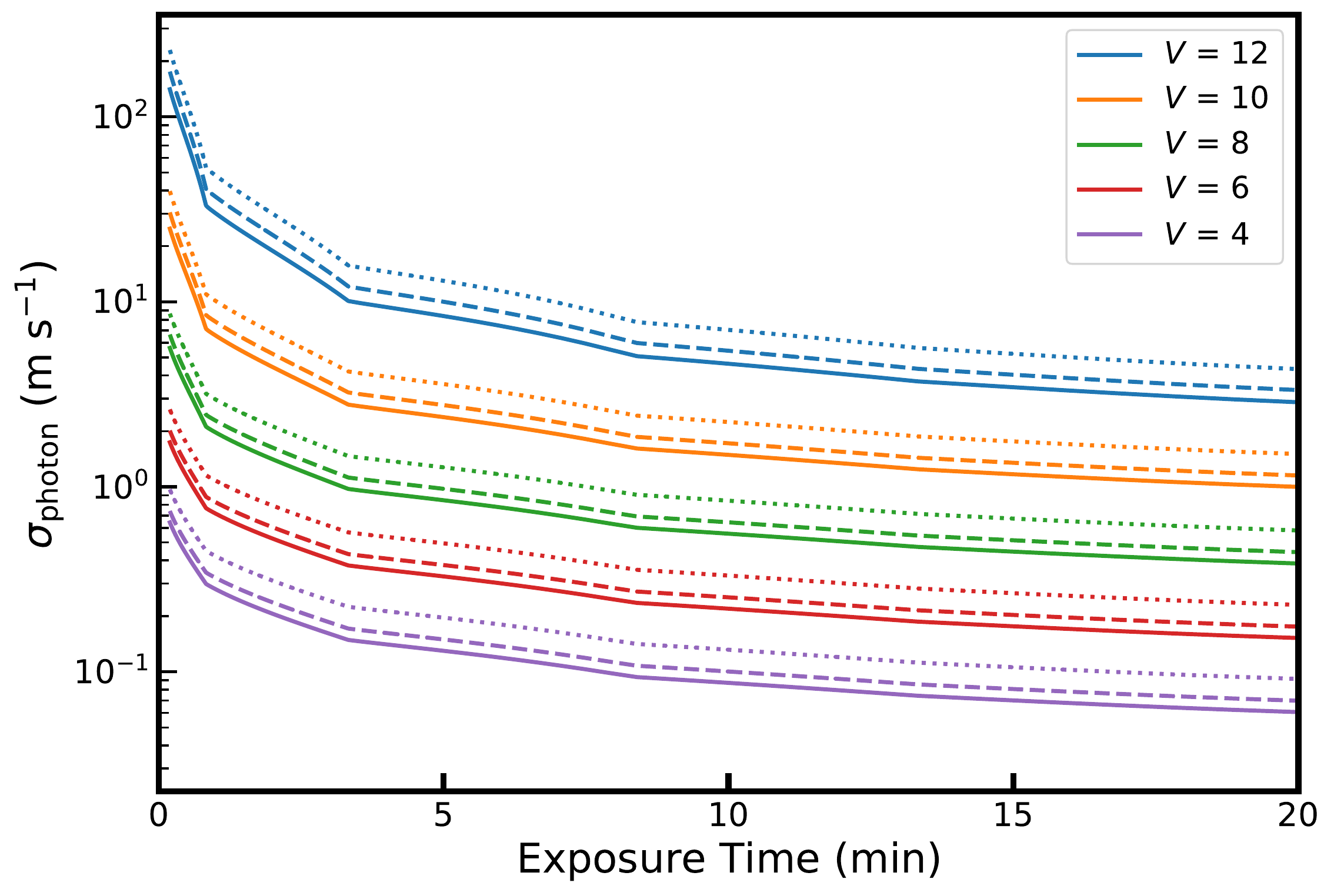}
    \caption{Predicted photon noise as a function of exposure time using the NEID exposure time calculator. For each of $V=$ 4 (purple), 6 (red), 8 (green), 10 (orange), and 12 (blue), we plot the photon noise for $\teff=$ 4500 K (solid line), 5500 K (dashed line), and 6500 K (dotted line). We assume median WIYN seeing ($0.8$") and airmass = 1.1.}
    \label{fig:calculator}
\end{figure*}

To reduce computation time for future use, we generate a grid of calculated RV uncertainties and SNR estimates spanning a range of stellar and observational parameters. The grid spans stellar effective temperatures from $2700$ K $\leq \teff \leq 6600$ K ($\log g = 4.5$ and solar abundance assumed), apparent \textit{V}-magnitudes $3 \leq V \leq 17$, and exposure times $10$ s $< t < 3600$ s. We assume airmass = 1.1 for the entire grid. Finally, a scaling factor is applied to account for the change in the fraction of starlight entering the NEID fiber under non-median seeing conditions, spanning a range of values between 0$\farcs$3 and 1$\farcs$9. Using an interpolation over this pre-computed grid in $\teff$, $V$, seeing, and exposure time, we have constructed an exposure time, SNR, and RV precision calculator\footnote{\url{http://neid-etc.tuc.noao.edu/calc_shell/about}} for NEID.
We show $\sigma_{\rm photon}$ as a function of exposure time for various $V$ and $\teff$ in Figure \ref{fig:calculator}.
Instrument commissioning is still in progress, but we note that the current performance exceeds the conservative exposure time calculator numbers given here.

\subsection{Stellar Noise}
\subsubsection{Stellar Rotation}

The broadening of spectral lines due to stellar rotation limits the precision with which RVs can be measured. The Doppler information content contained in the wings of spectral lines can be determined more precisely when the slope of the wings, i.e., the change in intensity with wavelength, is steeper \citep{Butler96}. Consequently, the measurement uncertainty scales proportionally with the projected stellar rotational velocity, $v\sin i$ \citep{Bouchy01}. While this is an intrinsic stellar effect, we do not treat it as an independent term in our error budget. Instead, stellar rotation is captured by the photon noise calculation by introducing a broadening kernel as discussed in the previous section.

\subsubsection{Convection}

\begin{figure}
    \epsscale{1.1}
    \centering
    \plotone{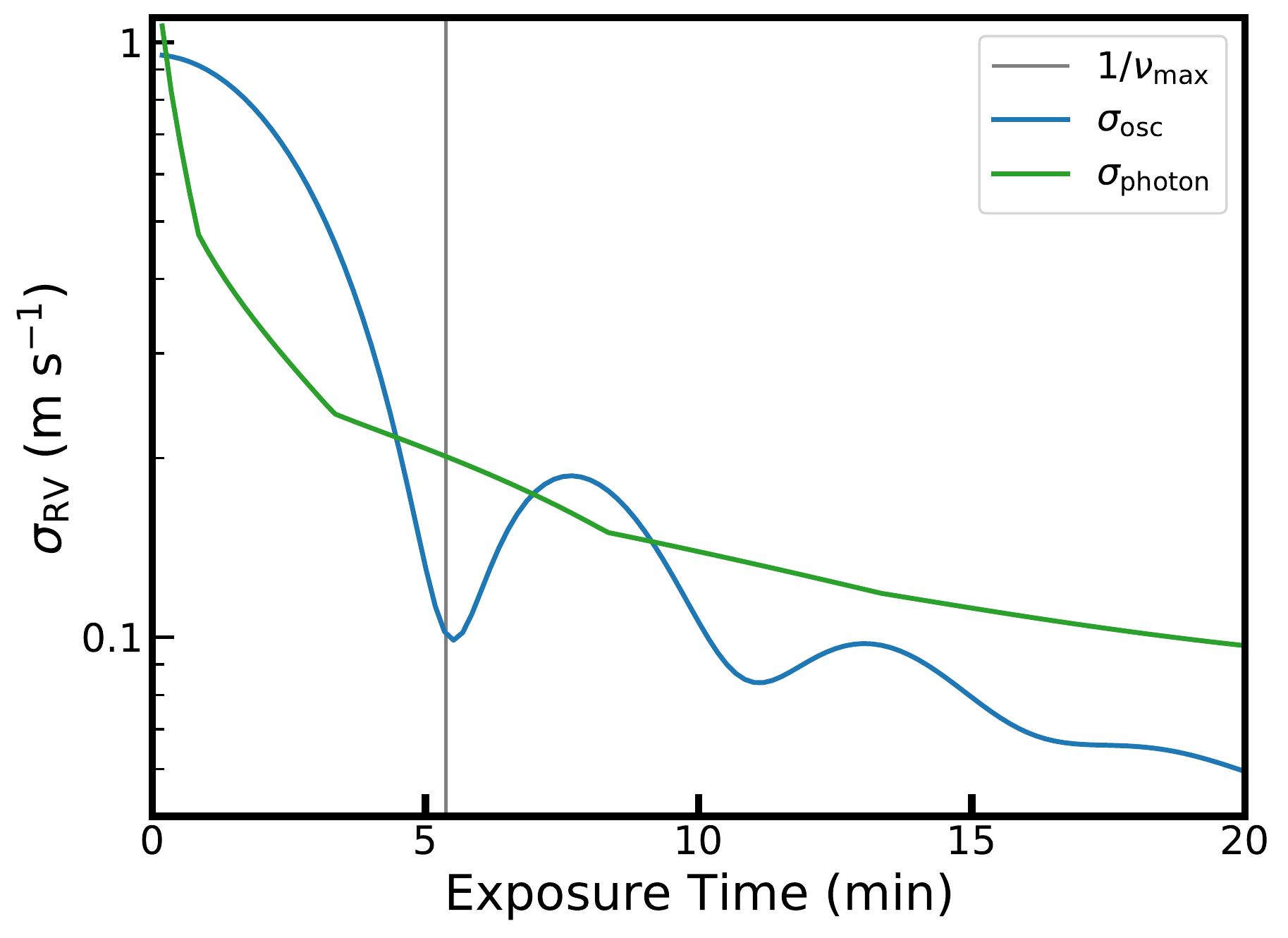}
    \caption{Residual p-mode amplitude (blue) as a function of exposure time for a Sun-like star based on the \citep{Chaplin19} model. We also plot the photon noise (green), assuming $V=5$, and we indicate the primary p-mode frequency of the Sun, $\nu_{{\rm max},\odot}=3100\upmu$Hz (vertical grey line). The residual p-mode amplitude reaches a trough at close to 10 \cms\ just after $t=1/\nu_{{\rm max},\odot}$.}
    \label{fig:psun}
\end{figure}

The RV signal imparted on a stellar spectrum by convective motions can be broken down into separate components due to surface granulation and p-mode oscillations. To estimate the average impact of granulation on our RV precision, we adopt a scaling relation from \citet{Kjeldsen11},
\begin{equation}\label{eq:gran}
    \sigma_{\rm gran} \propto \left(\frac{L_*}{L_\odot}\right)^{1/2}
    \left(\frac{M_*}{M_\odot}\right)^{-1}
    \left(\frac{T_{{\rm eff},*}}{T_{{\rm eff},\odot}}\right)^{-1/2},
\end{equation}
and we take the solar granulation RMS to be $\sigma_{{\rm gran},\odot}=$ 1 m s$^{-1}$ \citep{Luhn20}. The signal due to p-mode oscillations is similarly expected to be on the order of 1 m s$^{-1}$ \citep{Medina18, Yu18}, with the amplitude of the primary p-mode scaling as
\begin{equation}
    A_{\rm max} \propto \left(\frac{g_*}{g_\odot}\right)^{-1}
    \left(\frac{T_{{\rm eff},*}}{T_{{\rm eff},\odot}}\right)^{4}.
\end{equation}
But as \citet{Chaplin19} show, this signal can be filtered out on timescales as short as the duration of a single exposure, and the residual p-mode amplitude can be reduced to the 10 \cms\ level in just 5 - 10 minutes for Sun-like stars (Figure \ref{fig:psun}).

\begin{figure}
    \epsscale{1.05}
    \centering
    \plotone{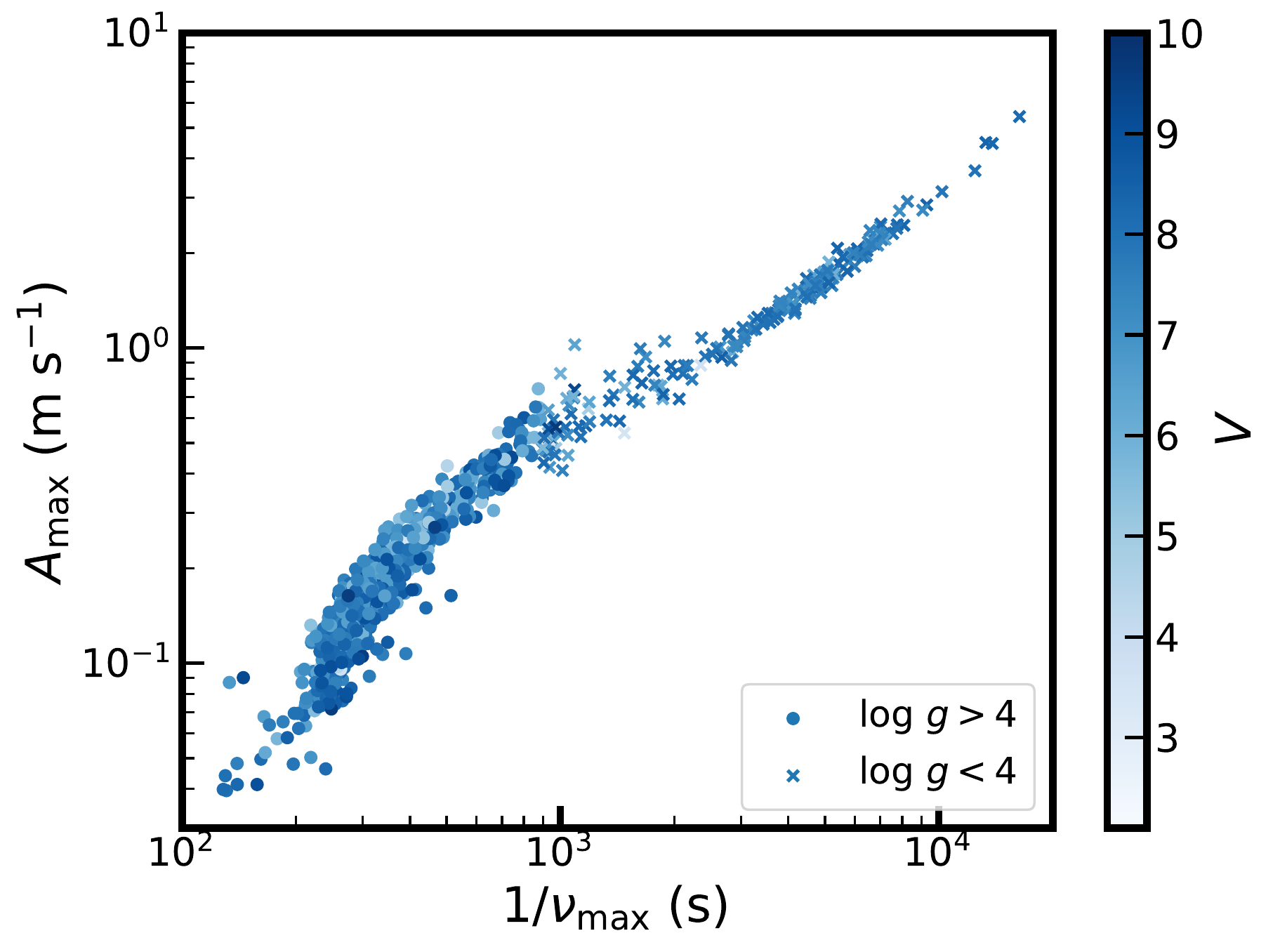}
    \caption{P-mode amplitudes and frequencies for all stars with spectroscopic $M_*$, $\teff$, log($g$), $v\sin i$, and log $R'_{\rm HK}$ from \citet{Brewer16}. Main sequence dwarfs, in the lower left, have low amplitudes that can be rapidly filtered, whereas subgiants and stars near the end of the main sequence, which occupy the upper right region of the diagram, have both large p-mode amplitudes and long filtering timescales, and are not conducive to precision RV observations.}
    \label{fig:pmodes}
\end{figure}

The characteristic timescale for filtering p-mode oscillations is given by  $\nu_{\rm max}$, the frequency of the strongest oscillation peak, which can be calculated using the theoretical scaling relation determined by \citet{Kjeldsen95}
\begin{equation}
    \nu_{\rm max} \propto \left(\frac{g_*}{g_\odot}\right)
    \left(\frac{T_{{\rm eff},*}}{T_{{\rm eff},\odot}}\right)^{-1/2}
\end{equation}
In Figure \ref{fig:pmodes}, we show the primary p-mode amplitude as a function of $\nu_{\rm max}$ for a subset of the \citet{Brewer16} catalog of F-, G-, and K-stars from the California Planet Search \citep{Howard10a}. Stars with strong p-modes typically also have long filtering timescales, making them poor candidates for RV exoplanet searches.

\subsubsection{Magnetic activity}\label{sec:magnetic}

Activity-induced features such as starspots and faculae suppress convective blueshift and modify the local stellar flux level, deforming spectral lines and altering the RV signal observed for a given star.
Because these features are rotationally-modulated and change quasi-periodically, they can not only mask \citep{Saar97} but also mimic the RV signal from an exoplanet \citep{Robertson15}. While activity can in principle be disentangled from exoplanet-induced signals via careful analysis of the RV time series \citep[e.g.,][]{Robertson14,Haywood14,Suarez2020}, this analysis often requires an impracticable number and frequency of observations. For blind exoplanet searches, magnetic activity is thus treated as a source of RV uncertainty.

A commonly used observational indicator for the level of stellar activity --- and the corresponding level of RV variability --- is the flux in the Ca H and K emission lines, as this flux is driven by chromospheric heating from magnetic fields.
In this work, we use the log $R'_{\rm HK}$ index \citep{Wright05}, which is a measurement of the flux in the Ca H and K line cores relative to the total flux in the $R$-band \citep{Noyes84}, to estimate the magnetic activity contribution to RV precision.

We explore the relation between RV variability and log $R'_{\rm HK}$ using the RV RMS values and median $S_{\rm HK}$ values measured by \citet{Luhn20} for a subset of stars observed with HIRES as part of the California Planet Search \citep{Howard10a}.
This stellar subset was selected by \citet{Luhn20} to satisfy the following criteria: $>10$ HIRES observations, $M_\star>0.7M_\odot$, and self-consistent spectroscopically-derived and isochrone-derived surface gravities as measured by \citet{Brewer16}.
We further restrict our analysis to stars with $\log g >4$ so that the residual p-mode amplitude will be at the sub-m s$^{-1}$ level.
For this work, we do not distinguish between variability arising from short-term rotational modulation and long-term magnetic activity cycles; our analysis reflects only the combined level of activity-induced uncertainty averaged over the HIRES observational time baseline.

\begin{figure*}
    \epsscale{1.0}
    \centering
    \plotone{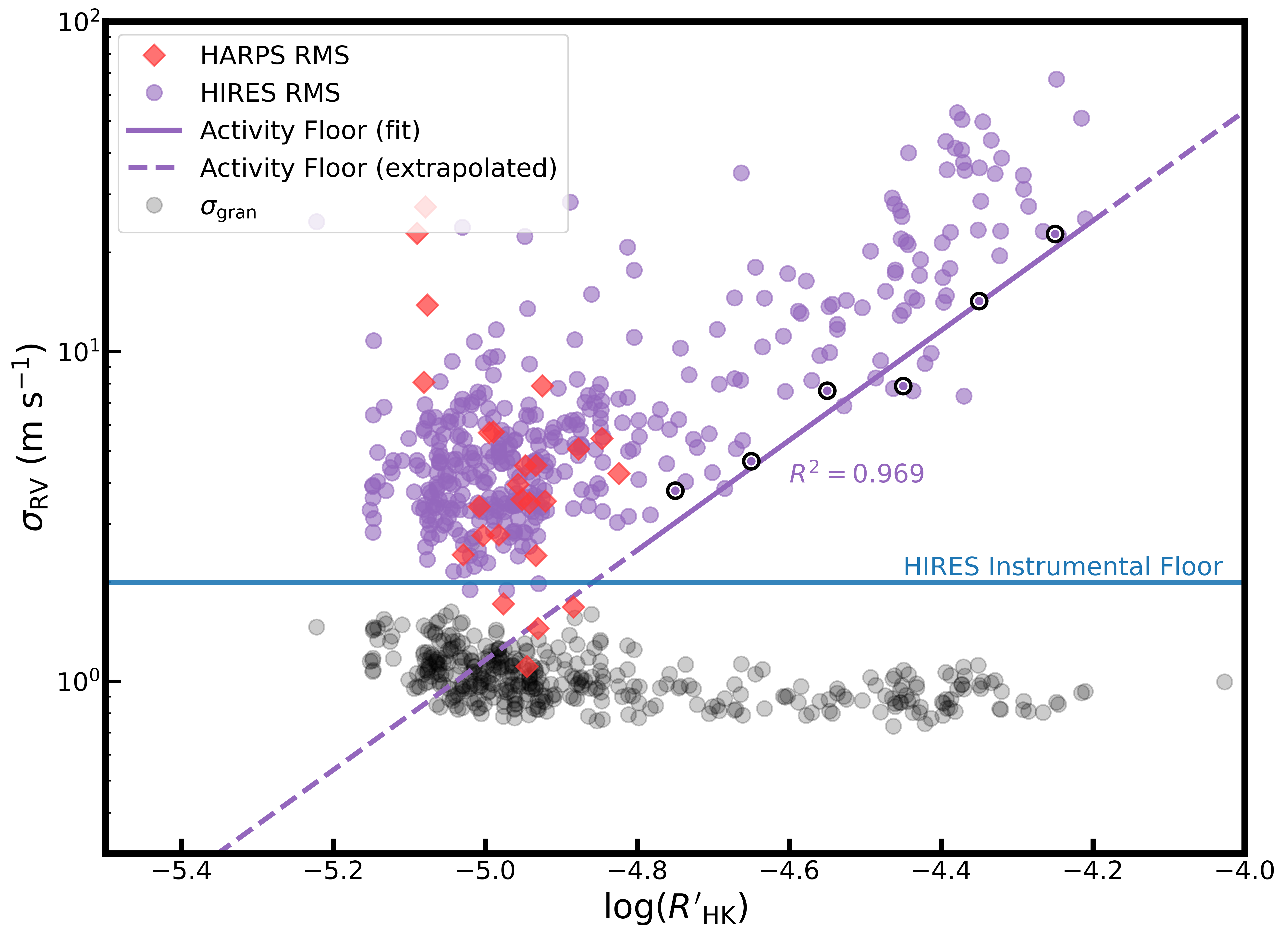}
    \caption{RV RMS as a function of $R'_{\rm HK}$ for a subset of California Planet Search stars. We plot HIRES RMS values (purple circles) as measured by \citet{Luhn20} and HARPS RMS values (red diamonds) as calculated in this work, excluding low surface gravity stars with $\log g<4$. The 5th percentile lower envelope points are outlined in black. We show the activity floor fit to data with $-4.8<\log R'_{\rm HK}<-4.2$ (solid purple line) and extrapolated to quieter stars (dashed purple line). We also plot $\sigma_{\rm gran}$ (black circles) for each star as calculated in Equation \ref{eq:gran}. The fit to the activity floor in the activity dominated regime appears to hold for lower values of $\log R'_{\rm HK}$, as indicated by the HARPS RMS values below the HIRES instrumental floor.}
    \label{fig:activity}
\end{figure*}

The median $\log R'_{\rm HK}$ index for each star is calculated from $S_{\rm HK}$ following \citet{Noyes84}.
As we show in Figure \ref{fig:activity}, the $\sigma_{\rm RV}$ values for stars with $\log R'_{\rm HK}>-4.8$ are activity-dominated, i.e., $\sigma_{\rm RV}$ is significantly larger than the expected uncertainty contributions from either granulation (calculated as in Equation \ref{eq:gran}) or oscillations.
In addition, there appears to be a lower envelope of $\sigma_{\rm RV}$, or an ``activity floor'', that varies with log $R'_{\rm HK}$ in this activity-dominated regime.
To derive an expression for this activity floor, we first divide the data into 0.1 dex bins from $\log R'_{\rm HK}=-4.8$ to $\log R'_{\rm HK}=-4.2$.
We then assume a constant instrumental uncertainty contribution of $2$ m s $^{-1}$ and subtract this out in quadrature from $\sigma_{\rm RV}$ to obtain the activity component of the RMS, $\sigma_{\rm mag} = \sqrt{\sigma_{\rm RV}^2-4}$.
The lower envelope is then calculated as the 5th percentile of the $\sigma_{\rm mag}$ data in each $\log R'_{\rm HK}$ bin.
A linear fit to the logarithm of this lower envelope yields
\begin{equation}\label{eq:mag}
    \log \sigma_{\rm mag}  = 1.66\log R'_{\rm HK} +8.39
\end{equation}
with a linear correlation coefficient of $R^2=0.969$.
Thus, in the activity-dominated regime, the activity floor exhibits a strong log-log correlation with $R'_{\rm HK}$.
That is, for a given level of stellar activity, the lowest activity-induced RMS we should expect to observe can be calculated as in Equation \ref{eq:mag}. 

To examine whether Equation \ref{eq:mag} holds true for quieter ($\log R'_{\rm HK}<-4.8$) stars, we dig beneath the $2$ m s $^{-1}$ HIRES instrumental floor using RV data from HARPS \citep{Mayor03}. To calculate the HARPS RMS, we take RV measurements from the HARPS pipeline for all stars that were observed by both HARPS and HIRES. We retain measurements with high SNR and individual uncertainties $<5$ m s$^{-1}$ and we remove outliers with RV values greater than 4-$\sigma$ from the mean. Observations taken on the same night are binned together and a systematic velocity offset is applied to data taken before the HARPS fiber upgrade \citep{LoCurto12}.
The HARPS RMS values are plotted in Figure \ref{fig:activity}, and we show that $\sigma_{\rm RV}$ does indeed continue to decrease with decreasing $\log R'_{\rm HK}$, though we note that contributions from granulation are no longer insignificant in this regime. If we extrapolate Equation \ref{eq:mag} to stars with $\log R'_{\rm HK} =-5$, we find that the activity floor drops to $\sigma_{\rm mag}\approx1$ m s$^{-1}$; this value is compatible with observations of solar activity with HARPS-N \citep{Dumusque20} and is representative of the level of activity-induced uncertainty for the quiet stars we expect to observe with NEID.

\subsection{Spectral Contamination}

Telluric features are introduced into observed spectra as starlight passes through the Earth's atmosphere en route to ground-based instruments such as NEID. Disentangling a stellar spectrum from that of the atmosphere is not a trivial task, particularly as the local atmospheric composition varies with time, so it is common practice to simply mask out and exclude strong telluric features when calculating radial velocities. However, at the sub-m s$^{-1}$ RV precision we expect to achieve with NEID, the uncertainty introduced by weaker telluric features, or `micro-tellurics', becomes significant \citep{Cunha14, Wang19}.

In addition to telluric contamination, observations conducted when the moon is up or during twilight will suffer from solar spectral contamination. The spectral features of the Sun will be offset relative to those of a target star when their relative barycentric radial velocities are nonzero. This will cause a peak-pulling effect in the measured CCF of the stellar spectrum, which can introduce RV uncertainties as large as $100$ m s$^{-1}$ \citep{Roy20}. This peak-pulling effect is strongest for relative radial velocities of $|\Delta$RV$|$ $\approx 4$ km s$^{-1}$, and it increases in strength for faint stars and bright sky backgrounds \citep{Roy20}.

Contributions to the single measurement precision from telluric and solar spectral contamination are folded into the $27$ \cms\ NEID instrumental error budget \citep{Halverson16}, in which it is determined that each of these signals can be calibrated out at the 10 \cms\ level. However, the suggested solar calibration techniques \citep{Halverson16,Roy20}, have yet to be validated with on-sky data. In \S\ref{sec:observations}, we explore the possibility of limiting the level of solar contamination via observation planning alone.

\section{NEID Earth Twin Survey Candidates}\label{sec:sample}
To identify the best candidates for NETS, we start with the sample of 166 G- and K-dwarfs from the Eta-Earth Survey \citep{Howard10b}. These bright, nearby stars were carefully selected to have properties conducive to precise RV measurement.
We note, however, that stars that failed to meet the search criteria of the Eta-Earth Survey may still prove to be promising exoplanet search targets. To explore this possibility, we also consider a larger sample of 1624 stars with Keck-HIRES RV data published by \citet{Butler17}.

As we discuss in \S\ref{sec:noise}, the total RV uncertainty for observations of a given star, can be predicted if the stellar properties are known\footnote{This will also depend on knowledge of the period and phase of the stellar activity cycle.}. The expected uncertainty is in turn an important factor in determining whether a star is a suitable exoplanet search target.
Owing to an extensive history of RV monitoring, the Eta-Earth targets are very well studied and their properties have been reliably characterized. This is not true of all stars in the latter sample, as observations of these have in many cases been sparse and sporadic.
We therefore restrict our analysis to only those stars for which we have strong constraints on $M_*$, $\teff$, log($g$), $v\sin i$, and log $R'_{\rm HK}$. For the F-, G-, and K-stars in our sample, we rely on spectroscopic parameter measurements from \citet{Brewer16}, and to extend the sample to lower masses, we also include the M-dwarfs GL 699 and GL 876 \citep{Mann15}, and HD 1326, HD 95735, HD 201091, and HD 201092 \citep{Boyajian12}. We then impose a declination limit of $\delta>-25^\circ$ and eliminate stars with significant rotational broadening ($v\sin i>5$ km s$^{-1}$), large p-mode amplitudes ($\log g<4$), and high magnetic activity indicators ($\log R'_{\rm HK}>-4.5$).
We analyze the remaining 431 stars according to the quantitative metrics defined in \S\ref{sec:selection} to construct a list of  targets for NETS.

Our cuts in $\log g$ and $\log R'_{\rm HK}$ allow for p-mode and activity contributions on the order of 10 m s$^{-1}$, which will dominate relative to the other sources of noise. But these cuts are simply intended as a means of removing the \textit{most} active and \textit{most} evolved stars. The best RV targets will ultimately be identified following our analysis in the next section. In addition, stars with large p-mode amplitudes and strong magnetic activity retain some value, as these signals can be accounted for via careful observation strategies \citep[e.g.,][]{Chaplin19} and analysis techniques \citep[e.g.,][]{Robertson14}.

\section{Target Selection Metrics}\label{sec:selection}

\subsection{Discovery Space}\label{sec:dspace}
Because the stars that we consider for our survey were all targets of past RV exoplanet searches, certain classes of planets have already been discovered for some systems and ruled out for others. One should not expect to discover hot Jupiters, for example, around stars from the Eta-Earth Survey, as these stars have already been thoroughly picked over for large RV signals. 
For a given star with $N$ RV observations at a single measurement precision of $\sigma_{\rm RV}$, \citet{Howard16} show that the minimum detectable RV semi-amplitude for an exoplanet of period $P$ can be calculated as
\begin{equation}\label{eq:detlim}
    K = \alpha \frac{\sigma_{\rm RV}}{\sqrt{N}}\sqrt{1+10^{(P/\tau-1.5)^2}},
\end{equation}
where $\tau$ is the time span over which the observations were taken. If the stellar mass, $M_*$, is known, one can then convert $K$ to $M_p\sin i$ 
\begin{equation}\label{eq:minmass}
    M_p\sin i \propto K (M_p+M_*)^{2/3}P^{1/3}(1-e^2)^{1/2}.
\end{equation}
The unitless scale factor $\alpha$ can be varied to adjust the probability of detecting a planet with semi-amplitude $K$ or minimum mass $M_p \sin i$. Decreasing $\alpha$ lowers the value of $M_p \sin i$ that Equation \ref{eq:minmass} returns, with the caveat that there is then a lower probability of detecting an exoplanet of the computed minimum mass. 
Using injection-recovery tests on archival data from the California Planet Search \citep{Howard10a}, \citet{Howard16} find that $\alpha = 6$ for a 50\% detection probability.

\begin{figure*}
    \epsscale{1.0}
    \plotone{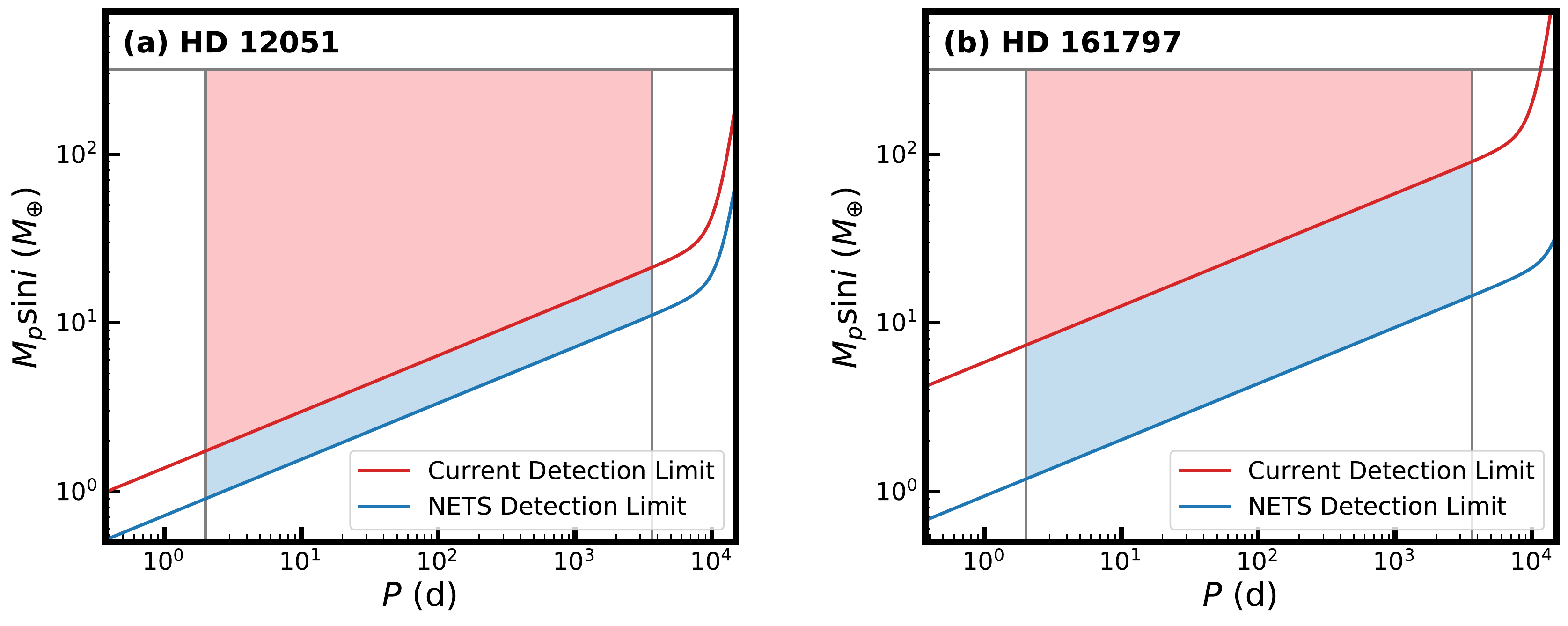}
    \caption{Predicted 50\% exoplanet detection limits for NETS (blue) and for RV observations to date (red).  Exoplanets in the red shaded region in $\log P$ - $\log M\sin i$ space should have been discovered by past surveys, while the blue shaded region marks the new discovery space, $\Delta$, that NETS will probe. We show an example of one star that has been thoroughly studied by past surveys (HD 12051, panel (a)), and one star that has been observed more sparsely (HD 161797, panel (b)). The grey lines denote the upper mass limit of $M_J$ and period bounds of 2 days $< P <$ 10 year used in Equation \ref{eq:dspace}.}
    \label{fig:dspace}
\end{figure*}

To quantify the potential for the discovery of new exoplanets, we define a ``discovery space'' metric, $\Delta$, as the integrated area in $\log P$ - $\log M_p \sin i$ space (Figure \ref{fig:dspace}) between detection limits from several past surveys and the improved limits we expect to achieve with NETS
\begin{equation}\label{eq:dspace}
    \Delta = 1.051\int_{2{\rm d}}^{10{\rm yr}}\log\left(\frac{[M_p\sin i]_{\rm current}}{[M_p\sin i]_{\rm NETS}}\right)d\log P
\end{equation}
where the normalization factor, $1.051$, is chosen such that $\Delta = 1$ for $\tau$ Ceti.
We restrict the range of periods to 2 days $< P <$ 10 years and set an upper mass limit of $M_p\sin i < M_J$.
We calculate detection limits using the empirical formulae of \citet{Howard16}, assuming that NEID will observe each GTO target 50 times each year for 5 years, and we take the single measurement precision to be $\sigma_{\rm RV} =  1.47$ m s$^{-1}$ using
\begin{equation}\label{eq:noise}
    \sigma_{\rm RV} = \sqrt{\sigma_{\rm inst}^2+\sigma_{\rm photon}^2+\sigma_{\rm osc}^2+\sigma_{\rm gran}^2+\sigma_{\rm mag}^2}
\end{equation}
where $\sigma_{\rm inst} = 27$ \cms, $\sqrt{\sigma_{\rm osc}^2+\sigma_{\rm photon}^2}$ is held at 30 \cms, and $\sigma_{\rm gran}$ and $\sigma_{\rm mag}$ are assumed to each be 1 m s$^{-1}$.  The minimum mass, $M_p \sin i$, is then computed as in Equation \ref{eq:minmass} with the simplifying approximation $M_*\gg M_p$ and assuming circular orbits ($e=0$). We choose $\alpha=6$ and a 50\% detection threshold when calculating our discovery space metric.
In S\ref{sec:altdspace}, we discuss the impact of our assumptions regarding the detection threshold, exploring its dependence on the completion fraction, and regarding uncertainty contributions from granulation and activity, which may be reduced to $\ll 1 $ m s$^{-1}$ by employing sophisticated mitigation techniques.

To determine the ``current'' detection limits for each target, we consider past observations from HIRES, HARPS, and SOPHIE \citep{Bouchy06}, and ongoing observations from HARPS-N. For HIRES, we take $N$, $\sigma_{\rm RV}$, and time baselines from \citet{Butler17}, and for HARPS, we use the values calculated in \S\ref{sec:magnetic}. To determine the corresponding numbers for SOPHIE, we analyze RV measurements retrieved from the SOPHIE archive\footnote{\url{http://atlas.obs-hp.fr/sophie/}}. We only use data collected after the octagonal fiber upgrade, after which the instrument was demonstrated to achieve precisions on par with those of HIRES and HARPS for bright standard stars \citep{Bouchy13}. As with the HARPS data, observations taken on the same night are binned together and we remove outliers with RV values greater than 4-$\sigma$ from the mean for each star. Observations that are still in their proprietary period as of this writing are excluded as well, because the archived RV values for these data are precise only to 100 m s$^{-1}$.  We assume that all stars in the HARPS-N GTO \citep{Motalebi15} will be observed 250 times at $\sigma_{\rm RV} = 2$ m s$^{-1}$ over the course of a 5 year survey.

\subsection{Exposure Time}
We can increase the total number of stars surveyed and augment the exoplanet discovery potential of the survey as a whole by prioritizing stars with shorter exposure time requirements.
To determine the typical exposure time for each potential target, we consider the short-timescale behavior of photon noise and p-mode oscillations. For the bright $V<8$ mag, high surface gravity stars we are considering for NETS, RV uncertainty contributions from both of these sources are expected to fall to sub-m s$^{-1}$ levels on similar 5 - 10 minute timescales.

We use the NEID exposure time calculator and the procedure outlined in \citet{Chaplin19} to calculate the minimum required exposure time, $t_{30}$, for which we can achieve a combined precision of $\sqrt{\sigma_{\rm osc}^2+\sigma_{\rm photon}^2} \leq30$ \cms. While we acknowledge that precisions as low as 10 \cms\ are attainable for both p-modes and photon noise, we stick to a 30 \cms\ threshold given that the impact on the total error budget will have a rapidly diminishing return as we probe uncertainty levels lower than the anticipated NEID instrumental floor. 

\begin{figure*}
    \centering
    \epsscale{1.0}
    \plotone{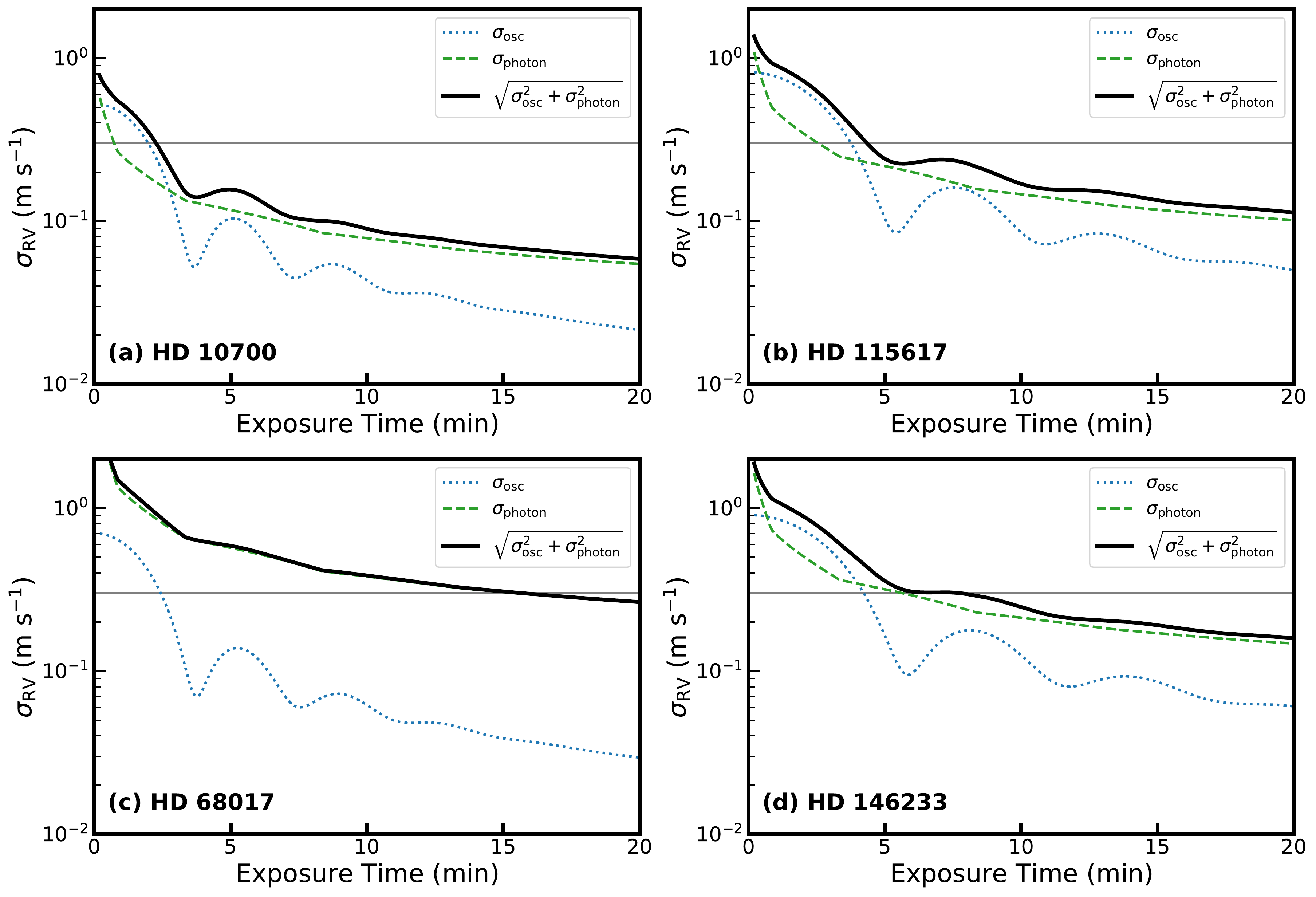}
    \caption{RV uncertainties introduced by photon noise and p-mode oscillations as a function of exposure time. P-modes are shown as a blue dotted line, photon noise is a green dashed line, and the combined contribution is shown as a solid black line. We also plot the 30 \cms\ threshold discussed in \S\ref{sec:selection}. In panels (a) and (b), we show $\sigma_{\rm RV}$ for the bright, low-mass stars HD 10700 ($\tau$ Ceti) and the sun-like star HD 115617, respectively. For these stars, we expect to reach the 30 \cms\ threshold before $t_p$, so we set $t_{\rm exp} = t_p$.  In panel (c), we show $\sigma_{\rm RV}$ for the faint star HD 68017, for which $t_{30}\gg t_p$. In this case, we set $t_{\rm exp} = t_{30}$.  Panel (d) shows $\sigma_{\rm RV}$ for HD 146233. It is evident that we achieve a minimal gain in precision when increasing the exposure time from $t=5$ minutes to $t=7$ minutes, and yet we do not reach the 30 \cms\ threshold until $t=7$ minutes. While it is true that this gain may not be worth the extended exposure time, we note that the residual p-mode amplitude curve will vary stochastically. Decreasing the exposure time may significantly elevate the noise, so we maintain $t_{\rm exp} = t_{30}$.}
    \label{fig:exptime}
\end{figure*}

For relatively bright stars, the photon noise drops off rapidly and we can reach our target 30 \cms\ precision with exposure times short enough that $t_{30} < t_p$, where $t_p = 1/\nu_{\rm max}$ is the primary p-mode period. But as we show in Figures \ref{fig:psun} and \ref{fig:exptime}, the residual p-mode amplitude decreases rapidly and approaches a local minimum as the exposure time approaches $t_p$. Increasing the exposure time beyond the 30 \cms\ threshold yields a significant gain in precision for relatively little extra cost. We therefore impose the additional requirement that each star is observed for at least one p-mode period.

We note that the first minimum in the residual p-mode amplitude curve is somewhat offset from $t_p$ given that the p-mode RV signal is constructed from multiple, simultaneously-excited modes beating together. However, due to the stochastic excitation and damping \citep{Chaplin19} and short lifetimes of individual p-modes \citep[several days for Sun-like stars,][]{Bedding04}, the interference pattern and total p-mode signal will vary in a manner that we cannot anticipate without nearly continuous RV monitoring of our targets. We therefore use $t_p$, a reasonable approximation of the average position of this first minimum, in defining our exposure time requirements. The second component of our figure of merit is thus
\begin{equation}\label{eq:texp}
    t_{\rm exp}=\left\{\begin{array}{ll}t_{30}, \ \  & t_p\leq t_{30} \\ t_p, \ \  & t_p>t_{30} \end{array}\right. .
\end{equation}

\subsection{Empirical RV RMS}\label{sec:emprms}
The timescales of p-mode oscillations are short enough that their contribution to stellar RV variability can be accounted for without a complex observing strategy. Granulation, however, manifests on timescales of hours to days \citep{Kjeldsen11}, and activity-induced signals vary with stellar rotation periods \citep{Saar97} and long-term activity cycles \citep{Lovis11}. While strategies for removing these sources of noise have been demonstrated \citep{Dumusque11}, it is nevertheless prudent to select stars for which the intrinsic level of variability is low.

To identify stars with low intrinsic RV variability, we examine the observed RV RMS of each of our potential targets as measured by \citet{Butler17} with long term trends and signals from known planets removed. The precision of these data are limited by the HIRES instrumental floor of roughly 2 m s$^{-1}$ \citep{Butler17}, however. So while we can use HIRES data to distinguish quiet stars from stars with high intrinsic variability, we are unable to easily differentiate between stars with empirical RMS values $\lesssim2$ m s$^{-1}$. We therefore use HARPS RMS values (see \S\ref{sec:magnetic}) in place of the HIRES numbers for stars that were observed by both instruments. Given that HARPS has a lower instrumental floor \citep{Fischer16}, we expect the HARPS RMS values to provide a better handle on the intrinsic stellar noise.

\subsection{Figure of Merit}\label{sec:fom}

Two final corrections are applied to the metrics listed above. First, we add a 180 s overhead to the exposure time metric to account for the estimated telescope slew and target acquisition time that will accompany each observation. We also note that the discovery space metric heavily favors stars with very few RV observations to date. While this certainly opens the door for the discovery of new planets, a lack of observations is also a sign that the empirical RMS value may be poorly constrained or unreliable. To balance these two factors, a logistic curve is used to weight the criteria in favor of stars with $N_{\rm obs}>10$ observations to date.

The candidates are then ranked using the following Figure of Merit (FOM)
\begin{equation}\label{eq:fom}
\begin{split}
    {\rm FOM}\ & = 0.5\log(\Delta) - 0.5\log(1+e^{10-N_{\rm obs}}) \\
    & -\log({\rm RMS}) - \log(t_{\rm exp}+180 {\rm s}) + k
\end{split}
\end{equation}
where $\Delta$ is calculated as in Equation \ref{eq:dspace}, $t_{\rm exp}$ is calculated as in Equation \ref{eq:texp}, and the RMS is described in \S\ref{sec:emprms}. The coefficients in this equation reflect the perceived importance of each term, and $k = 7.5$ is an arbitrary, constant offset chosen so that $\rm{FOM} > 0$ for all stars. The top 100 stars were selected as NETS targets, and the Figure of Merit, discovery space, exposure time, RMS, and number of observations for each of these stars are given in Table \ref{tab:startable}. 

\startlongtable
\begin{deluxetable*}{rlrrrrrrrr}
\tablecaption{Candidates for the NEID Earth Twin Survey\label{tab:startable}}
\tablehead{
\colhead{Rank} &
\colhead{Star Name} & \colhead{$V$} &
\colhead{$\teff$} & \colhead{$\log g$} & 
\colhead{$\Delta$} & \colhead{$t_{\rm exp}$} & 
\colhead{RMS} & \colhead{$N_{\rm obs}$} &\colhead{FOM} \\
\colhead{} &
\colhead{} & \colhead{} &
\colhead{(K)} & \colhead{} & 
\colhead{} & \colhead{(s)} & 
\colhead{(m s$^{-1}$)} & \colhead{} & \colhead{}
} 
\startdata
1&HD172051&5.85&5636&4.58&1.41&448&1.11&21&4.732\\
2&HD4614&3.46&5919&4.37&5.35&383&2.80&45&4.666\\
3&HD219834B&5.20&5135&4.48&3.69&277&3.40&63&4.592\\
4&HD143761&5.40&5833&4.29&5.35&457&3.00&186&4.582\\
5&HD86728&5.40&5742&4.31&4.02&433&3.20&48&4.509\\
6&HD10700&3.49&5333&4.60&1.00&214&2.50&272&4.506\\
7&HD201091&5.21&4361&4.63&0.76&180&2.40&84&4.504\\
8&HD190360&5.70&5549&4.29&3.66&446&3.20&143&4.480\\
9&HD10780&5.63&5344&4.54&3.34&380&3.43&15&4.479\\
10&HD157214&5.38&5817&4.61&2.70&378&3.10&44&4.478\\
11&HD186427&6.20&5747&4.37&4.81&710&2.70&118&4.460\\
12&HD4628&5.74&4937&4.54&0.64&352&1.72&37&4.440\\
13&HD185144&4.67&5242&4.56&0.55&233&2.20&227&4.412\\
14&HD34411&4.70&5873&4.26&1.88&492&2.70&83&4.378\\
15&HD179957&6.75&5741&4.42&3.10&1016&2.00&97&4.367\\
16&HD10476&5.24&5190&4.51&0.72&260&2.80&133&4.336\\
17&HD186408&5.96&5778&4.28&3.53&806&2.90&86&4.317\\
18&HD38858&5.97&5735&4.46&1.32&546&2.41&117&4.317\\
19&HD50692&5.76&5913&4.39&2.60&642&3.20&81&4.287\\
20&HD68017&6.78&5626&4.60&4.53&942&3.10&105&4.286\\
21&HD115617&4.74&5562&4.44&1.41&316&3.96&228&4.281\\
22&HD154345&6.76&5455&4.52&3.52&864&3.00&141&4.278\\
23&HD55575&5.55&5888&4.32&2.24&722&2.80&55&4.273\\
24&HD52711&5.93&5886&4.39&2.62&660&3.30&78&4.266\\
25&HD19373&4.05&5938&4.18&2.14&595&3.30&85&4.258\\
26&HD95735&7.49&3464&4.86&1.72&640&2.80&165&4.256\\
27&HD217107&6.18&5575&4.25&6.21&844&4.35&123&4.248\\
28&HD221354&6.76&5221&4.47&0.96&796&1.80&171&4.247\\
29&HD51419&6.94&5732&4.51&2.59&1130&2.20&92&4.247\\
30&HD196761&6.36&5473&4.55&2.17&644&3.52&36&4.206\\
31&HD26965&4.43&5092&4.51&2.93&257&7.87&99&4.196\\
32&HD146233&5.50&5785&4.41&1.53&538&3.47&161&4.196\\
33&HD182572&5.17&5587&4.15&2.31&618&3.91&83&4.187\\
34&HD24496&6.81&5531&4.50&4.89&936&4.20&96&4.174\\
35&HD110897&5.95&5911&4.49&2.72&566&4.70&44&4.173\\
36&HD126053&6.25&5714&4.54&2.13&660&3.80&80&4.160\\
37&HD116442&7.06&5155&4.54&2.13&982&2.80&56&4.152\\
38&HD9407&6.52&5672&4.45&0.72&826&1.90&206&4.146\\
39&HD187923&6.16&5774&4.23&1.78&944&2.79&77&4.128\\
40&HD168009&6.30&5808&4.33&3.54&808&4.50&34&4.126\\
41&HD48682&5.25&6039&4.27&2.63&862&3.70&43&4.124\\
42&HD124292&7.05&5458&4.48&2.55&1094&3.00&35&4.121\\
43&HD185414&6.73&5848&4.49&6.54&1046&5.00&60&4.120\\
44&HD145958B&7.50&5343&4.46&5.34&1634&3.10&77&4.114\\
45&HD99491&6.49&5438&4.40&2.68&724&4.50&125&4.105\\
46&HD12846&6.89&5700&4.48&2.08&1078&2.90&69&4.097\\
47&HD164922&6.99&5341&4.39&2.38&1000&3.40&158&4.085\\
48&HD157347&6.30&5712&4.42&1.47&708&3.56&87&4.084\\
49&HD149806&7.10&5275&4.42&3.58&1086&4.00&34&4.072\\
50&HD170657&6.81&5040&4.54&3.19&778&5.15&43&4.058\\
51&HD199960&6.21&5885&4.22&3.27&1032&4.16&25&4.055\\
52&HD158633&6.44&5256&4.58&0.71&626&3.00&40&4.042\\
53&HD166620&6.38&4970&4.51&0.71&536&3.40&67&4.040\\
54&HD39881&6.60&5770&4.39&2.92&966&4.39&45&4.031\\
55&HD176377&6.80&5877&4.52&2.40&1110&3.60&87&4.023\\
56&HD33021&6.15&5786&4.21&2.78&994&4.26&23&4.023\\
57&HD30708&6.78&5707&4.26&3.37&1212&3.98&16&4.021\\
58&HD43947&6.61&5963&4.33&2.81&1164&3.80&39&4.016\\
59&HD3651&5.88&5221&4.45&0.77&442&4.30&92&4.016\\
60&HD1461&6.46&5739&4.34&0.87&854&2.77&251&4.012\\
61&HD69830&5.95&5387&4.48&1.42&466&5.70&268&4.009\\
62&HD221356&6.50&5987&4.31&3.12&1154&4.11&30&4.008\\
63&HD56303&7.34&5917&4.30&2.77&2032&2.37&16&4.002\\
64&HD1326&8.07&3567&4.89&1.52&1220&2.80&170&3.998\\
65&HD12051&7.14&5412&4.38&0.97&1206&2.32&232&3.986\\
66&HD58781&7.24&5582&4.37&2.77&1440&3.39&23&3.981\\
67&HD44985&7.01&5983&4.36&3.38&1462&3.71&17&3.980\\
68&HD18803&6.62&5648&4.45&2.47&874&5.00&113&3.975\\
69&HD191785&7.34&5132&4.48&2.48&1296&3.61&56&3.970\\
70&HD92719&6.79&5796&4.47&2.46&1066&4.27&66&3.970\\
71&HD84737&5.10&5872&4.10&2.56&1300&3.70&46&3.965\\
72&HD218868&7.00&5505&4.42&2.98&1074&4.80&78&3.958\\
73&HD135101A&6.69&5677&4.20&2.93&1122&4.65&42&3.952\\
74&HD37008&7.74&4979&4.53&2.11&1820&2.58&45&3.949\\
75&HD34575&7.09&5551&4.25&2.34&1336&3.63&23&3.944\\
76&HD84117&4.93&6128&4.26&3.24&902&6.04&55&3.940\\
77&HD71479&7.17&5925&4.27&3.26&1790&3.33&16&3.940\\
78&HD116443&7.35&5001&4.57&2.16&1266&3.71&92&3.938\\
79&HD197076&6.45&5810&4.42&2.12&872&5.07&42&3.936\\
80&HD151541&7.56&5309&4.50&1.81&1710&2.62&77&3.934\\
81&HD190404&7.28&4960&4.59&2.91&1178&4.78&54&3.920\\
82&HD132142&7.77&5145&4.55&1.96&1952&2.55&58&3.911\\
83&HD42250&7.43&5357&4.43&3.24&1536&4.08&25&3.910\\
84&HD159222&6.56&5870&4.41&2.86&976&5.80&88&3.902\\
85&HD98281&7.29&5381&4.52&2.55&1338&4.17&68&3.901\\
86&HD144585&6.32&5807&4.23&3.68&996&6.49&35&3.900\\
87&HD145958A&7.43&5414&4.48&5.19&1564&5.20&84&3.900\\
88&HD125455&7.58&5103&4.54&2.65&1606&3.65&45&3.898\\
89&HD9562&5.75&5837&4.02&3.00&1678&3.78&30&3.892\\
90&HD62613&6.55&5493&4.50&0.76&758&3.80&30&3.889\\
91&HD130992&7.81&4767&4.51&2.36&1846&3.17&51&3.879\\
92&HD219623&5.58&6059&4.21&4.21&1062&6.92&20&3.878\\
93&HD32923&5.00&5725&4.13&2.56&655&8.08&21&3.875\\
94&HD180161&7.04&5396&4.51&4.45&1046&7.30&14&3.872\\
95&HD4915&6.98&5668&4.56&2.72&1118&5.45&40&3.868\\
96&HD220339&7.80&4953&4.56&2.32&1908&3.15&53&3.865\\
97&HD224619&7.47&5453&4.49&3.05&1666&4.09&33&3.864\\
98&HD28005&6.72&5747&4.24&2.65&1234&4.98&57&3.864\\
99&HD37394&6.21&5249&4.50&5.27&498&14.66&13&3.864\\
100&HD216520&7.53&5082&4.54&1.44&1522&3.08&185&3.860\\
\enddata
\end{deluxetable*}

Given the NETS time allocation, we can realistically expect to observe just 25-35 stars for the duration of the survey. While we will largely select candidates based on their rankings, the selection process will also allow for some subjectivity to capture criteria that are not encompassed in the Figure of Merit. The following considerations will be taken into account:
\begin{itemize}
    \item Observability: We must choose targets with an appropriate on-sky distribution such that observations can be spread out over the course of each year.
    \item Diversity of Stellar Types: The current understanding of stellar variability suggests that G and K main sequence stars are the best targets for RV exoplanet searches. But with the advent of NEID and other EPRV spectrographs, we expect our understanding of stellar RV noise at the sub-m s$^{-1}$ level to evolve. We aim to select stars from a variety of stellar types for NETS so that we are not hampered in the event that stellar noise does not adhere to the current picture.
    \item Known Exoplanets: While the discovery space metric captures the potential for detecting new exoplanets, it does not consider exoplanets that have already been discovered. These systems offer the potential for studies of multi-planet systems and may merit additional consideration. We will also consider the effects of known planets on the dynamical stability of orbits in the circumstellar habitable zone.
\end{itemize}

\subsection{Alternate Discovery Space Approximations}\label{sec:altdspace}

\begin{figure*}
    \centering
    \epsscale{1.0}
    \plotone{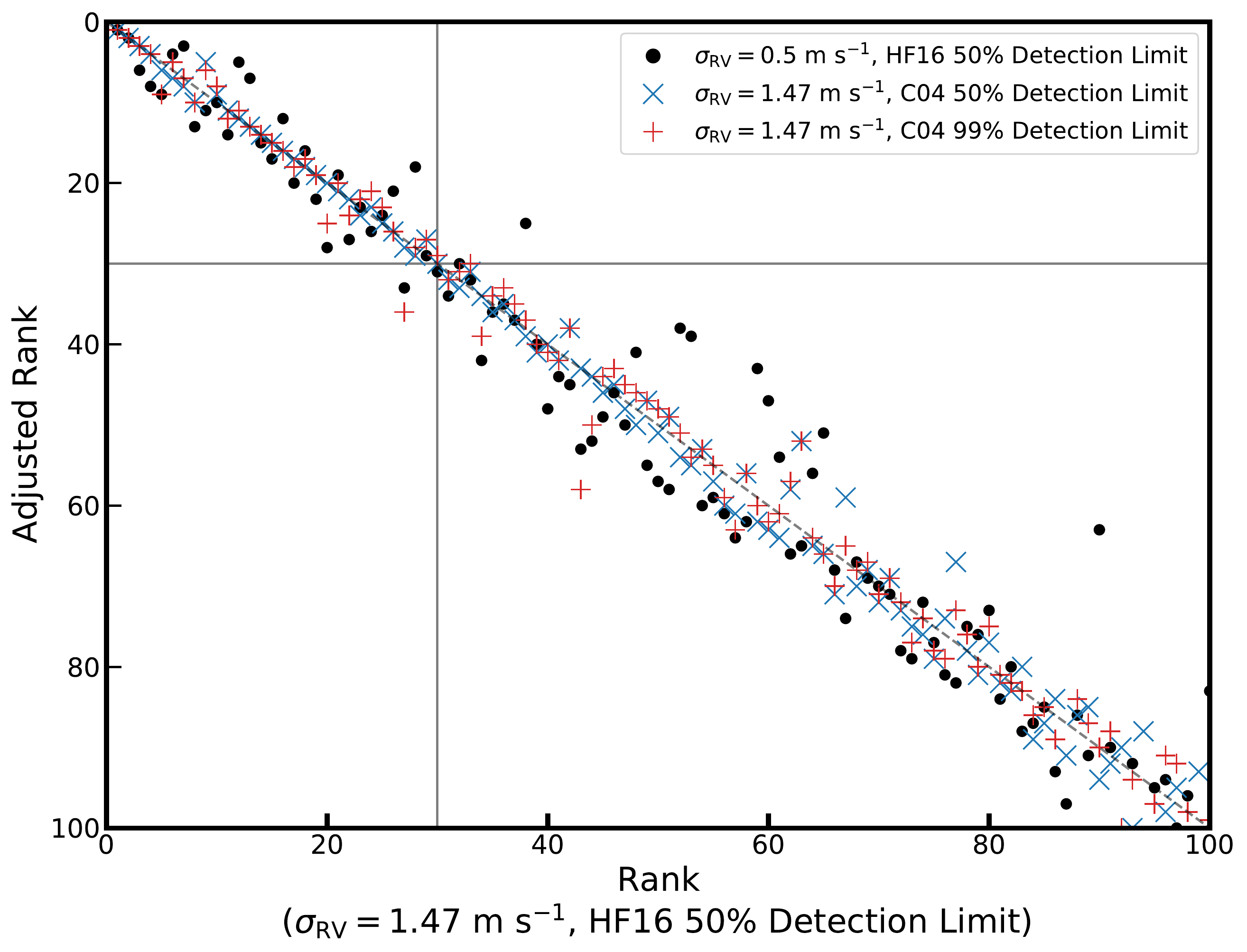}
    \caption{Adjusted target rankings for the initial NETS target stars under the various approximations described in \S\ref{sec:altdspace}. We show the effect of reducing the RV uncertainty contributions of granulation and stellar activity (black circles) and of using the \citet{Cumming04} (C04) detection limit approximations with $\epsilon_D=0.5$ (blue Xs) and $\epsilon_D=0.99$ (red plus signs) in place of the empirical limits determined by \citet{Howard16} (HF16). Very few stars move in to or out of the upper left region of the plot, which contains the highest priority targets.}
    \label{fig:rank_compare}
\end{figure*}

In \S\ref{sec:dspace}, we make several assumptions when calculating our discovery space metric, $\Delta$. Here, we consider the impact of these assumptions on the final rankings we calculate in \S\ref{sec:fom}. First, we determined the expected single measurement precision for NEID observations under the assumption that granulation and stellar activity will each contribute at the 1 m s$^{-1}$ level. These uncertainties are significant relative to photon noise, p-mode oscillations, and the instrumental precision floor, and they make up the majority of the total error budget. But as mitigation techniques improve in the near future, it is conceivable that uncertainty contributions from activity and granulation will be limited to the 30 cm s$^{-1}$ level as well. In recent years, significant progress has been made in identifying the signs of stellar activity and granulation \citep{Hojjatpanah20,Wise18,Davis17,Meunier17,Meunier19a,Thompson17} and in calibrating these signals out \citep{Rajpaul15,Langellier20,Meunier19b,ForemanMackey17,Dumusque18,Zhao20,Gilbertson20,Haywood20,Milbourne19,Cretignier20}. To explore how our target prioritization should change in this optimistic scenario, we re-compute the Figure of Merit for each of our targets using a NETS single measurement precision of $\sigma_{\rm RV}=0.5$ m s$^{-1}$ in the discovery space metric. Figure \ref{fig:rank_compare} compares the adjusted rankings to those given in Table \ref{tab:startable}.

The mean change in rank for the top 100 stars is $\pm4.76$, but this change should have little effect on the final NETS target list. If we take the list to be comprised of the 30 highest ranked stars, then stars in the upper left section of Figure \ref{fig:rank_compare} will make the cut and be included in the survey sample regardless of our assumptions about the NETS single measurement precision, whereas those in the lower right will fail to make the cut in both cases. Only the four stars that fall in the upper right and lower left will be added to and removed from the sample, respectively, if we use the adjusted rankings to construct our list. These are the stars for which the improved precision and detection limits lead to a relatively large (upper right) or small (lower left) increase in the anticipated discovery space.

We also base the discovery space metric in \S\ref{sec:dspace} off of a 50\% detection threshold, but the search completeness of the survey at, e.g., the 99\% level may be of more interest in certain cases. We redo the analysis in \S\ref{sec:dspace} and calculate adjusted rankings using the detection limit approximations given by \citet{Cumming04}
\begin{equation}
    K=\left\{\begin{array}{ll}\frac{2 K_{0}}{1-\cos (\pi \tau / 2 P)}, \ \  & \tau<P<\frac{\pi T}{8\left(1-\epsilon_{\mathrm{D}}\right)} \\
\frac{K_{0}}{2\left(1-\epsilon_{\mathrm{D}}\right)}\left(\frac{P}{\pi \tau}\right), \ \  & P>\frac{\pi T}{8\left(1-\epsilon_{\mathrm{D}}\right)}
\end{array}\right.
\end{equation}
where $\epsilon_D$ is the detection probability of interest. We show the adjusted rankings for $\epsilon_D=0.5$ and $\epsilon_D=0.99$ in Figure \ref{fig:rank_compare}. The mean difference between the rankings computed using the \citet{Howard16} 50\% and \citet{Cumming04} 50\% thresholds is $\pm2.00$ and between the \citet{Howard16} 50\% and \citet{Cumming04} 99\% thresholds is $\pm2.45$. Again taking the final target list to be comprised of the top 30 stars, we show that varying the detection threshold does not impact our target selection.

\section{Observational Limitations on Radial Velocity Precision}\label{sec:observations}
\subsection{Charge Transfer Inefficiency}

The Charge Transfer Efficiency (CTE) of a CCD describes the fractional efficiency with which electrons are moved from pixel to pixel during detector readout; conversely, the Charge Transfer \emph{In}efficiency (CTI = 1 - CTE) quantifies the cumulative charge lost \citep{Goudfrooij06}. Due to imperfections in the detector lattice, CTI will in practice always be nonzero. This can be a concern for RV exoplanet searches, as the flux deficit due to CTI translates to an offset in the measured RV relative to a perfectly efficient detector \citep{Bouchy09}. Because we are most interested in the precision of the time variability of the measured RV signal rather than the accuracy of its absolute value, a constant offset alone is not a significant obstacle. But the CTI accumulated during readout (and the corresponding RV offset) is a function of the total collected flux, or SNR, which in turn may change by more than a factor of two for exposures of fixed duration due to reasonable variations in seeing, airmass, and transparency. As \citet{Blake17} show using a detector similar to the NEID CCD, the offset can vary in excess of 30 \cms\ for $\Delta$CTI$>10^{-7}$.

To limit the uncertainty contribution from CTI to the 5 \cms\ allowed by the NEID instrumental error budget \citep{Halverson16}, NETS exposures will be set to trigger on a constant target SNR for all observations of a given star. The target SNR for each star will correspond to the exposure time we calculate in \S\ref{sec:selection}. This will ensure that the observations satisfy our photon noise and p-mode uncertainty requirements and that the observational cost we estimate here will be close to the real cost incurred in typical observing conditions.

\subsection{Detector Saturation and Nonlinearity}
For the target selection criteria in \S\ref{sec:selection}, we require observations to be no shorter than the star's primary p-mode period (Eq. \ref{eq:texp}). This requirement is easily satisfied for faint stars, for which exposures will not hit the target SNR until well after a single p-mode period has elapsed. For several very bright stars, however, we find that the detector reaches nonlinearity or even saturates on timescales shorter than $t_p$. Observations of these stars will be split into multiple exposures, with the target SNR for each exposure held constant but the number of exposures per observation varying with observing conditions so that the total observing time spans the p-mode period. In this way, we can still average out p-mode signals without saturating the detector.

\subsection{Solar Contamination}
\citet{Roy20} outline several techniques for calibrating out the RV uncertainties introduced by peak pulling due to solar spectral contamination. They also show, however, that the $\sigma_{\rm RV}$ contribution can be limited to $< 10$ \cms\ (the maximum level permitted by the \citet{Halverson16} instrumental uncertainty budget) by observing stars with relative barycentric radial velocities $|\Delta \rm{RV}|\gg4$ km s$^{-1}$ and/or stars that are $>12$ magnitudes brighter than the solar contribution to the sky background.
Here we explore the contributions to the dark sky background from the two sources of solar contamination with which we are concerned: scattered sunlight during twilight and scattered moonlight.

\begin{figure*}
    \centering
    \epsscale{1.0}
    \plotone{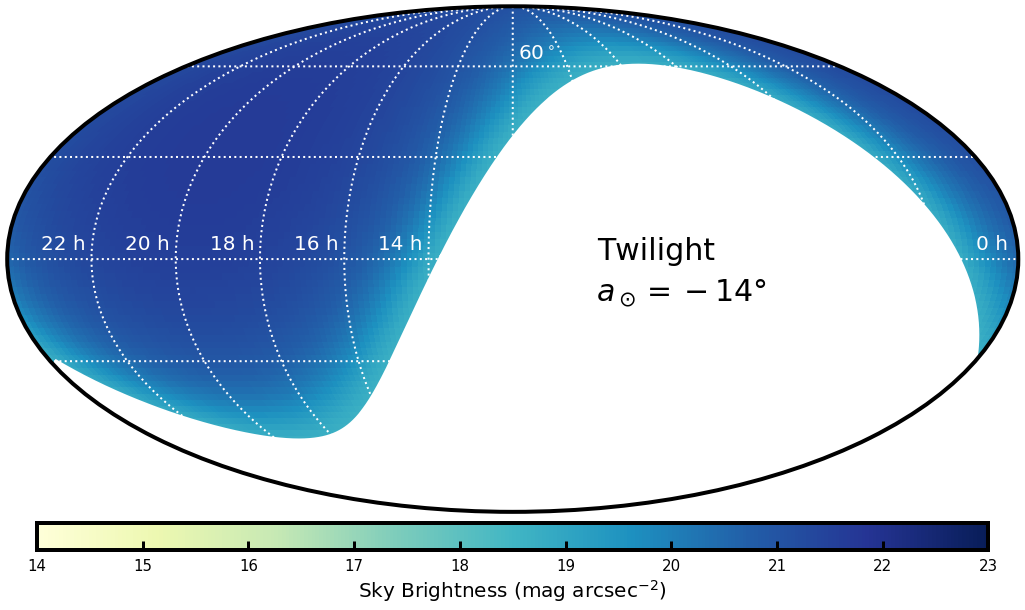}
    \caption{Scattered sunlight contribution to the sky brightness ($V$ mag arcsec $^{-2}$) when the sun is $14^\circ$ below the horizon.}
    \label{fig:twi_bright}
\end{figure*}

We model the twilight contribution to the background sky brightness as in \citet{Yoachim16}, substituting a dark time zenith sky brightness of $V = 21.95$ mag arcsec $^{-2}$ for Kitt Peak \citep{Neugent10}. This zenith sky brightness value was shown to remain constant from 1988 to 2009 \citep{Massey00,Neugent10}, so we do not expect that it has changed significantly over the past decade. We show the solar contribution to the sky brightness as observed from Kitt Peak between nautical and astronomical twilight ($a_\odot=-14^\circ$) in Figure \ref{fig:twi_bright}.
For typical NETS targets ($V<8$ mag), twilight contamination will only be a serious concern if the stars are observed when they are close to the horizon.

\begin{figure*}[h]
    \centering
    \epsscale{1.0}
    \plotone{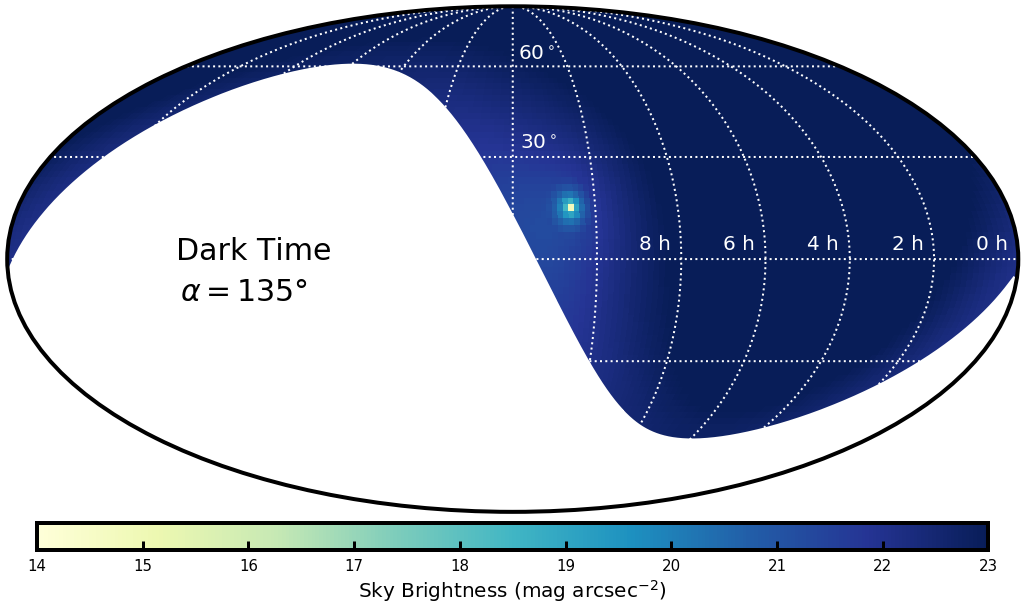}
    \caption{Scattered moonlight contribution to the sky brightness ($V$ mag arcsec $^{-2}$) for dark time ($\alpha=135^\circ$, where $0^\circ\leq\alpha\leq180^\circ$ and $\alpha=0^\circ$ corresponds to a full moon). The scattered moonlight contribution is insignificant across most of the sky, but rapidly increases within $30^\circ$ of the moon.}
    \label{fig:moon_dark}
\end{figure*}

\begin{figure*}[h]
    \centering
    \epsscale{1.0}
    \plotone{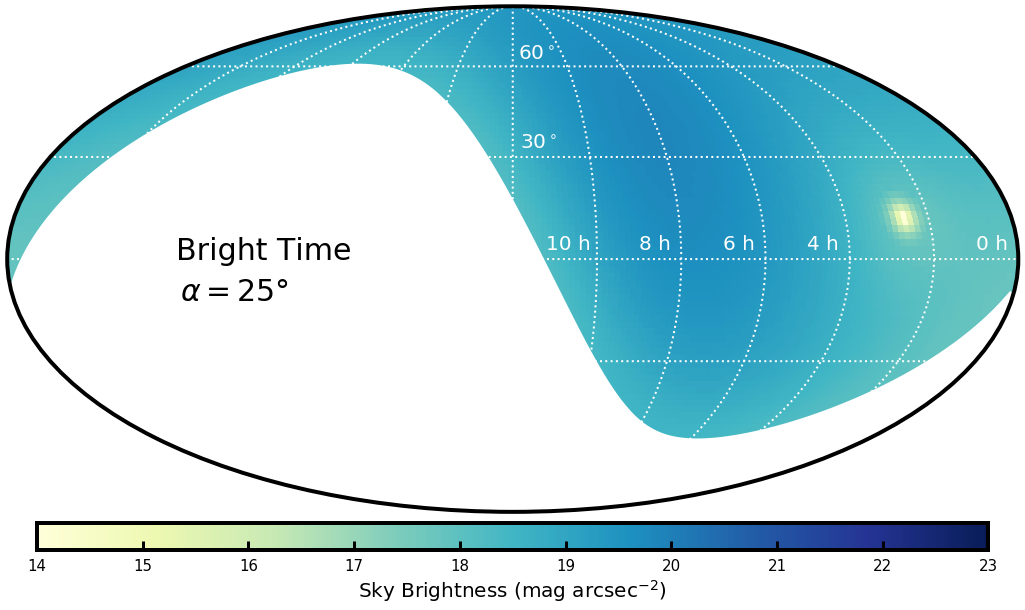}
    \caption{Same as Figure \ref{fig:moon_bright} but for bright time ($\alpha=25^\circ$). The scattered moonlight contribution is brighter than $20$ mag arcsec $^{-2}$ across the sky.}
    \label{fig:moon_bright}
\end{figure*}

For scattered moonlight, we rely on the model of \citet{Krisciunas91}, again taking $V = 21.95$ mag arcsec $^{-2}$ \citep{Neugent10} to be the dark time zenith sky brightness. This contribution is shown in Figures \ref{fig:moon_dark} and \ref{fig:moon_bright} for dark time and bright time, respectively. Scattered moonlight is not expected to contribute at a level $>10$ \cms\ during dark time except in the immediate vicinity of the moon. During bright time, the RV uncertainty contribution may be significant for any star fainter than $V = 6$ across most of the sky. 

These sky background models, in conjunction with the results of \citet{Roy20} and knowledge of the difference between the barycentric radial velocities of the Sun and the star being observed, will be used to predict the expected uncertainty contributions from solar contamination. We can take advantage of the flexible, queue-based NEID observing schedule and ensure that our targets are observed only when contamination is weak.


\section{Discussion and Conclusion}

In this work, we present an overview of NETS, a search for Earth-mass exoplanets around the nearest and brightest Solar-type stars. In \S\ref{sec:noise}, we describe an exposure time calculator for NEID with which the photon noise contribution to the RV precision can be assessed. We emphasize, however, that the numbers reported by the current version of the calculator are preliminary, pending completion of instrument commissioning. 
We expect our estimates of both $\sigma_{\rm inst}$ and $\sigma_{\rm photon}$ to evolve during commissioning as instrument subsystems are tested and as calibration algorithms are validated with on-sky data.
We also provide a brief, but comprehensive, overview of the external sources of RV uncertainty, namely intrinsic stellar variability and spectral contamination. We describe the expected level of uncertainty due to stellar granulation and oscillations and we derive an expression for the stellar ``activity floor'' as a function of the $R'_{\rm HK}$ index, which places a lower limit on the magnetic activity contribution to the total RV uncertainty. We acknowledge, however, that the observed level of variation for a given star can be higher or lower depending on where the star lies in its activity cycle, an effect that is not considered in our analysis. Ongoing studies of activity indicators over long timescales (e.g., Baum et al., in prep.) may inform target selection by enabling the identification of stars that are close to their activity minimum.

In \S\ref{sec:selection}, we outline a set of quantitative selection metrics that can be used to evaluate stars for consideration in RV exoplanet searches, and we use these metrics to construct an initial NETS target list. In defining the selection metrics, we discuss a strategy for jointly mitigating photon noise and p-mode RV uncertainties for observations of bright stars based on the work of \citep{Chaplin19}. We show that with careful planning, p-mode uncertainties can be accounted for with relatively little extra observational cost. We also consider the potential for breakthroughs in mitigation of granulation and activity, and while these will certainly influence the results of our survey, we show that there is no significant impact on our target prioritization scheme.

We highlight considerations for detector saturation and CTI variations and delve deeper into the issue of solar contamination in \S\ref{sec:observations}. We present a model to predict the total solar contribution to the background sky brightness and, together with the work of \citet{Roy20}, the associated RV uncertainty contribution. This model will be an essential observation planning tool as we search for sub-m s$^{-1}$ signals with NEID. We plan to add functionality to the NEID exposure time calculator to assess uncertainty contributions from both p-mode oscillations and solar spectral contamination and to provide observing recommendations to account for theses effects. In addition, with appropriate changes to normalization constants, our solar contamination model can easily be used to plan observations with instruments at other locations.

While the selection metrics and observing strategies in \S\ref{sec:selection} and \S\ref{sec:observations} are discussed in the context of NEID and NETS, these are broadly applicable to RV observations with any high-precision spectrograph.
Indeed, as constraints on stellar uncertainties tighten and as instrumentation and analysis techniques improve, the methodology described here can be extended to inform future surveys targeting RV precisions of 10 \cms\ and lower.
For surveys that remain limited by available observing resources, it will be prudent to weigh the relative observational costs of potential targets. As we illustrate in the context of p-mode oscillations, these costs must account not only for per-exposure photon noise but also for timescales of stellar variability and the appropriate observing schemes for uncertainty mitigation.
And while surveys that replicate past results certainly have merit, new exoplanet discoveries will always carry a higher priority. Target selection decisions should therefore take advantage of available records of previous work and de-prioritize stars for which new discoveries will be few and far between. We capture this information succinctly in our discovery space metric, which can trivially be adapted to new target lists and archival data sets.

Once a final NETS target list has been selected, we can explore the expected survey yield by simulating sets of NEID observations for each target with realistic cadences and baselines set by seasonal and nightly observing windows. Using these simulated observations, the true discovery space and \citep[given appropriate assumptions regarding occurrence rates, e.g.,][]{Hsu19} exoplanet yield can be predicted. We note, however, that these predictions will depend strongly on assumptions regarding the level of intrinsic stellar variability for each star and the mitigation strategies that we expect to employ, both of which are significant unknowns at this time. As such, we anticipate that these simulations and yield estimates will be of limited use in informing the survey strategy for NETS. But once NETS and other EPRV surveys are well underway and mitigation techniques at the sub-m s$^{-1}$ have been fleshed out, results of this nature can be used to guide the design of future RV instruments and surveys.

\acknowledgements

A.F.G.\ would like to thank Zhao Guo and Bill Chaplin for helpful discussions regarding p-mode mitigation.
NEID is funded by NASA through JPL by contract 1547612. We acknowledge support from the Heising-Simons Foundation via grant 2019-1177.
The Center for Exoplanets and Habitable Worlds and the Penn State Extraterrestrial Intelligence Center are supported by the Pennsylvania State University and the Eberly College of Science.
This research has made use of the SIMBAD database, operated at CDS, Strasbourg, France, and NASA's Astrophysics Data System Bibliographic Services.
Based on data retrieved from the SOPHIE archive at Observatoire de Haute-Provence (OHP), available at atlas.obs-hp.fr/sophie.

\software{Astropy \citep{astropy}, Matplotlib \citep{matplotlib}, NumPy \citep{numpy}, SciPy \citep{scipy}}




\end{document}